\documentclass[aps,prb,twocolumn,superscriptaddress,amsmath,amssymb,preprintnumbers,showpacs]{revtex4}

\usepackage{graphicx,color}
\usepackage{bm}
\newcommand{\bra}[1]{\langle{#1}|}
\newcommand{\ket}[1]{|{#1}\rangle}

\definecolor{dgreen}{rgb}{0,0.5,0}

\begin{document}


\title{Diffusion and transfer of entanglement in an array of inductively coupled flux qubits}

\author{Rosanna Migliore}\email{rosanna@fisica.unipa.it}
\affiliation{MIUR, CNISM, CNR-INFM, and Dipartimento di Scienze Fisiche ed Astronomiche, Universit\`a di Palermo, Via Archirafi 36, I-90123 Palermo, Italy}

\author{Kazuya Yuasa}
\email{kazuya.yuasa@ba.infn.it}
\affiliation{Dipartimento di Fisica, Universit\`a di Bari, I-70126 Bari, Italy}
\affiliation{Istituto Nazionale di Fisica Nucleare, Sezione di Bari, I-70126 Bari, Italy}

\author{Marina Guccione}\email{guccione@fisica.unipa.it}
\affiliation{MIUR, CNISM, and Dipartimento di Scienze Fisiche ed Astronomiche, Universit\`a di Palermo, Via Archirafi 36, I-90123 Palermo, Italy}

\author{Hiromichi Nakazato}\email{hiromici@waseda.jp}
\affiliation{Department of Physics, Waseda University, Tokyo 169-8555, Japan}

\author{Antonino Messina}\email{messina@fisica.unipa.it}
\affiliation{MIUR, CNISM, and Dipartimento di Scienze Fisiche ed Astronomiche, Universit\`a di Palermo, Via Archirafi 36, I-90123 Palermo, Italy}
\date[]{July 10, 2007}
\begin{abstract}
A theoretical scheme to generate multipartite entangled states in
a Josephson planar-designed architecture is reported. This scheme
improves the one published in [Phys.\ Rev.\ B \textbf{74},
104503 (2006)] since it speeds up the generation of W entangled
states in an $M\times N$ array of inductively coupled Josephson flux
qubits by reducing the number of necessary steps. In addition, the same protocol is shown to be able to transfer the W state from one row to the other.
\end{abstract}
\maketitle

\section{Introduction}\label{sec1}
In the last years, condensed-matter architectures based on
Josephson-junction qubits have appeared to be promising candidates
for quantum information processors.\cite{makhlin,noritoday} These
solid-state systems can be scaled up to a large number of qubits and
satisfy DiVincenzo's requirements for quantum
computing,\cite{DiVincenzo} i.e.\ state preparation, manipulation,
and readout. Among them, qubits on the basis of the superconducting
quantum interference devices (SQUID) offer the possibility of
realizing switchable (inductive) interbit couplings
\cite{castellano,plourde,norideph} and therefore provide promising
platforms to generate multipartite entanglements among
``macroscopic'' quantum systems in deterministic ways. Remarkable
experimental achievements with flux qubits include the realization
of complex single-qubit operation schemes,\cite{[3]} the generation
of entangled states \cite{[4],[5]} in systems of coupled flux
qubits,\cite{[3a]} while the direct measurement of the entanglement
in Josephson architectures has been performed via state tomography
\cite{martinis_scienze} for two superconducting phase qubits.

Within such frameworks, we proposed a scheme for the generation of a
W entangled state in a chain of $N$ spatially separated flux qubits
by exploiting their sequential couplings with one of them playing
the role of an entanglement mediator.\cite{noiPR2006} We remind that
the $N$-partite W entangled state is a natural generalization to
$N$-qubit state of the tripartite W state
$|W\rangle_3=(|100\rangle+|010\rangle+|001\rangle)/\sqrt3$, i.e.,
the totally symmetric (apart from possible phase factors) quantum
superposition of $N$ two-state systems where only one of them is in
its excited state. In Ref.\ \onlinecite{noiPR2006}, the success of
this scheme relies on the possibilities of both preparing the
initial state of the qubits and tuning the coupling energy and/or
the interaction time between each qubit and the mediator, provided
the time necessary for the desired quantum processes is short enough
with respect to the decoherence time.

The protection against noise is evidently one of the central issues
in quantum information technology and the reduction of the duration
spent for specific quantum operations is important for it. In this
paper, we improve the scheme proposed in Ref.\
\onlinecite{noiPR2006} by analyzing the dynamics of an
\textit{array} of flux qubits, which can be selectively coupled
in pairs for instance by exploiting the tunable flux transformer
proposed by Castellano \textit{et~al.},\cite{castellano} and
demonstrate how it is possible to \textit{diffuse} a W state
prepared in one row to two or more rows with a few steps. Such a
scheme helps to reduce the time for the generation of multipartite
entanglement. Furthermore, we show that the same protocol also
provides us with a way to shift, or \textit{transfer}, the W state
from row to row. We emphasize that W states are promising
candidates for the experimental realization of quantum information
processing in multiparticle systems since they possess entanglement
robustness against local operation even under qubit loss.

\section{An Array of Qubits}
\begin{figure}
\includegraphics[width=0.3\textwidth]{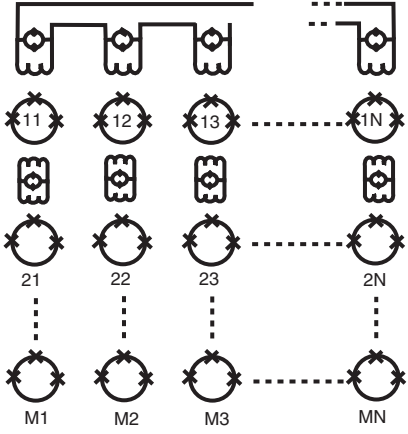}
\caption{Schematic illustration of an $M\times N$ array of Josephson
flux qubits. An inductive qubit-qubit coupling is realized by means
of a superconducting switch, namely a transformer with variable
flux-transfer function $\mathcal{R}(\phi_{cx}^{(mn)})$, as
proposed in Ref.\ \onlinecite{castellano}. It is possible to control
the flux-transfer function, and therefore the inductive-coupling
constant, by modulating the critical current of the inner dc-SQUID
of each transformer via an externally applied magnetic flux
$\phi_{cx}^{(mn)}$.
Each individual coupling between a pair of qubits is effectively turned on by adjusting the control fluxes of the
relevant ``switches'' with all the other ``switches'' kept off.
}\label{array}
\end{figure}

The idea presented in this paper is based on the theoretical
proposal in Ref.\ \onlinecite{noiPR2006} to generate a W state in a
chain of rf-SQUID or persistent current (3JJ) qubits. We first
recapitulate its essential idea within the present setup illustrated
in Fig.\ \ref{array}. Here, in order to minimize the susceptibility
to external noise of a large-inductance rf-SQUID, as proposed by
Mooij \textit{et~al.},\cite{mooij99} we consider the planar array
sketched in  Fig.~1 constituted by $M\times N$ spatially separated,
and consequently not directly interacting, tunable 3JJ qubits, that
is a superconducting loop containing three Josephson junctions, two
of equal size (i.e.\ with $E_{J,1}=E_{J,2}=E_J$) and the third one
smaller by a factor $\alpha$ (i.e.\ with $E_{J,3}=\alpha E_J$,
$\alpha<1$). This parameter may be adjusted for instance by
substituting the third junction with a dc-SQUID behaving as an
effective JJ with tunable Josephson energy, $E_{J,3}\equiv
E_{J,3}(\phi_c)$, $\phi_c$ being an additional control flux
threading the dc-SQUID loop. In such conditions, by applying an
external flux $\phi_x$ close to a half-integer number of flux
quanta, $\phi_0=h/2e$, and choosing $\alpha \approx 0.8$ the
potential energy of the total system forms a double well which
permits two stable configurations of minimum energy corresponding to
two persistent currents  $\pm I_p\approx\pm2\pi\alpha E_J/\phi_0$ in
the loop. This fact allows to engineer a two-state quantum system
(qubit) whose effective Hamiltonian, in the basis of the two energy
eigenstates $\ket{0_{mn}}$ and $\ket{1_{mn}}$ of the $(m,n)$ qubit
of the array (which at $\phi_x=\phi_0/2$, are maximal superpositions
of the two persistent-current states $\ket{L_{mn}}$ and
$\ket{R_{mn}}$), reads
\begin{equation}\label{hsqui}
H_{mn}=\frac{1}{2}\hbar\omega_{mn}
\sigma_z^{(mn)},\ \ %
\omega_{mn} =\sqrt{\Delta_{mn}^2(\phi_c)+\epsilon_{mn}^2(\phi_x)},
\end{equation}
where $m=1,\ldots, M$, $n=1,\ldots, N$, and
$\sigma_z^{(mn)}=\ket{1_{mn}}\bra{1_{mn}}-\ket{0_{mn}}\bra{0_{mn}}$
is a Pauli operator for the $(m,n)$th qubit. The energy spacing,
corresponding to a transition frequency $\omega_{mn}$ typically in
the range of microwaves, can be tuned by properly selecting both the
tunneling frequency, $\Delta_{mn}$, between $\ket{L_{mn}}$ and
$\ket{R_{mn}}$  and
$\epsilon_\text{mn}(\phi_x)=2I_p(\phi_x-\phi_0/2)/\hbar$ both
depending on the system parameters.


In the following discussion, we assume that all the qubits have a
common energy gap $\omega=\omega_{mn}$, $\forall(m,n)$.

The qubits are coupled with each other as depicted in Fig.~\ref{array}.
The coupled dynamics of the total system is described by the Hamiltonian
\begin{equation}\label{reduced}
H=H_0+\sum_{m,k=1}^M\sum_{n,\ell=1}^NH_{mn,k\ell}',\quad
H_0=\sum_{m=1}^M\sum_{n=1}^NH_{mn},
\end{equation}
where
\begin{equation}\label{RWA}
H_{mn,k\ell}'=g_{mn,k\ell}(\sigma_+^{(mn)}\sigma_-^{(k\ell)}
+\sigma_-^{(mn)}\sigma_+^{(k\ell)})
\end{equation}
is the rotating-wave coupling between qubits $(m,n)$ and $(k,\ell)$
with $\sigma_+^{(mn)}=\ket{1_{mn}}\bra{0_{mn}}$ and
$\sigma_-^{(mn)}=\ket{0_{mn}}\bra{1_{mn}}$ the raising and lowering
operators for qubit $(m,n)$, correspondingly.

The coupling constants
$g_{11,1n}=2(\xi_{11,1n}/L)(\phi_0^2/2)$
($n=2,\ldots,N$) between the first and the $n$th qubit in the first
row can be turned on and off via controlling the magnetic fluxes
$\phi_{cx}^{(mn)}$ externally applied to the $(m,n)$ qubit as shown
in Fig.~\ref{array}. Analogously we may control the coupling
constants
$g_{mn,(m+1)n}=2(\xi_{mn,(m+1)n}/L)(\phi_0^2/2)$ (with
$m=1,\ldots,M-1$, $n=1,\ldots,N$) between the $m$th and the
($m+1$)th qubits in the  $n$th column. Here
$\xi_{11,1n}=r_{11}\mathcal{R}(\phi_{cx}^{(11)})\mathcal{R}(\phi_{cx}^{(1n)})r_{1n}$
and
$\xi_{mn,(m+1)n}=r_{mn}\mathcal{R}(\phi_{cx}^{(m+1)n)})r_{(m+1)n}$,
$r_{mn}$ being the flux transforming ratio between the arm of the
transformer and the qubit $(m,n)$. In the following calculations,
we work in the interaction picture with respect to $H_0$.

If the inter-row couplings are turned off, $g_{mn,(m+1)n}=0$, the system is essentially the one analyzed in Ref.\ \onlinecite{noiPR2006}.
It is therefore possible to generate an $N$-partite W state in the first row, among qubits $(1,1),\ldots,(1,N)$, as follows:
\begin{itemize}
\item We prepare the initial state $\ket{\Psi_0}=\ket{1_{11}\ldots0_{1N}}\otimes\ket{0_{21}\ldots0_{2N}}\otimes\cdots\otimes\ket{0_{M1}\ldots0_{MN}}$,
with only the qubit $(1,1)$ in the excited state and the rest of
the qubits of the array in their own ground state.
\item The coupling $g_{11,12}$ is turned on during a proper time interval $0<t<\tau_1$ with other couplings off.
\item The coupling $g_{11,12}$ is turned off at $t=\tau_1$ and instead $g_{11,13}$ is on during $\tau_1<t<\tau_1+\tau_2$.
\item In this way, qubit $(1,1)$ is coupled with $(1,2),\ldots,(1,N)$ one by one.
\end{itemize}
It is shown in Ref.\ \onlinecite{noiPR2006} that, by exploiting
the knowledge of the coupling constants  $g_{11,1(n+1)}$ when the
relevant interaction is turned on, by setting  the interaction times
$\tau_n$ so that
\begin{equation}
\label{eqn:TuningN}
\sin\theta_n=\frac{1}{\sqrt{N-n+1}},\quad
\cos\theta_n=\sqrt{\frac{N-n}{N-n+1}},
\end{equation}
with $\theta_n=g_{11,1(n+1)}\tau_n/\hbar$ ($n=1,\ldots,N-1$), a W
state
\begin{equation}
\ket{w_1}
=\ket{W}_1\otimes\ket{0_{21}0_{22}\ldots0_{2N}}
\otimes\cdots\otimes\ket{0_{M1}0_{M2}\ldots0_{MN}}
\label{initial}
\end{equation}
with
\begin{align}
\ket{W}_m
=\frac{1}{\sqrt{N}}\,\Bigl(
\ket{1_{m1}0_{m2}\ldots0_{mN}}
&-i\ket{0_{m1}1_{m2}\ldots0_{mN}}\nonumber\\
-\cdots&-i\ket{0_{m1}0_{m2}\ldots1_{mN}}
\Bigr)
\label{wM}
\end{align}
is generated in the first row $m=1$ of the array.

We underline that rapid-single-flux-quantum (RSFQ)
Josephson-junction based logic circuits\cite{chiarello2,RSFQ} make
it possible to produce flux pulses characterized by rise/fall times
$t_{r/f}$ of the order of $10\,\text{ps}$. Therefore we are able to
obtain switching times much smaller than the duration of any step in
our scheme, typically less than or of the order of the inverse of
coupling energy $\hbar/g_j\approx 2\,\text{ns}$.

\section{Diffusion and Transfer of W State}\label{sec3}
Let us now discuss schemes for spreading the W state (\ref{initial}) prepared in the first row of the array to other rows, as well as for transferring it from row to row, by making use of the switchable inductive coupling $g_{mn,(m+1)n}$ between each qubit of the $m$th row ($m=1,\ldots,M-1$) and the corresponding one in the $(m+1)$th row.
To this end, we consider the following ``collective'' step by step scheme:
\begin{itemize}
\item  Each of the $N$ qubits in the first row, already prepared in the W state \eqref{initial}, is put in inductive interaction with the corresponding one in the second row during $0<t<\tau_\text{I}$, by turning on the couplings $g_{11,21}=g_{12,22}=\cdots=g_{1N,2N}=g_\text{I} \neq 0$, while other qubits evolve freely.
\item At $t=\tau_\text{I}$, the couplings $g_{11,21},g_{12,22},\ldots,g_{1N,2N}$ are turned off, and $g_{21,31}=g_{22,32}=\cdots=g_{2N,3N}=g_\text{II} \neq 0$ are turned on for time interval $\tau_\text{I}<t<\tau_\text{I}+\tau_\text{II}$, in order to couple the qubits in the second row with the corresponding ones in the third.
\item Similarly, the interactions between adjacent rows are successively switched on and off.
\end{itemize}
Then, by properly selecting the interaction times (by turning on and
off the coupling constants), it is possible to transfer the W state
from row to row or to diffuse it to multi-rows. To illustrate the
mechanism, let us look at a $3\times3$ array ($M=N=3$) for the sake
of simplicity. In this case, the W state prepared in the first row,
Eq.\ (\ref{initial}), reads
\begin{equation}
\ket{w_1}=\ket{W}_1
\otimes\ket{0_{21}0_{22}0_{23}}
\otimes\ket{0_{31}0_{32}0_{33}}.
\label{initial3}
\end{equation}
By switching on the inductive couplings $g_{11,21}=g_{12,22}=g_{13,23}=g_\text{I}\neq 0$ for a time period $\tau_\text{I}$, the W state $\ket{w_1}$ in (\ref{initial3}) is driven into
\begin{align}\label{1step}
\ket{\Psi_\text{I}}
=\frac{1}{\sqrt{3}}\,\Bigl(
\cos\theta_\text{I}
\ket{1_{11}0_{12}0_{13}}&\otimes\ket{0_{21}0_{22}0_{23}}
\nonumber\displaybreak[0]\\[-1mm]
{}-i\cos\theta_\text{I}
\ket{0_{11}1_{12}0_{13}}&\otimes\ket{0_{21}0_{22}0_{23}}
\nonumber\displaybreak[0]\\[1mm]
{}-i\cos\theta_\text{I}
\ket{0_{11}0_{12}1_{13}}&\otimes\ket{0_{21}0_{22}0_{23}}
\nonumber\displaybreak[0]\\[1mm]
{}-i\sin\theta_\text{I}
\ket{0_{11}0_{12}0_{13}}&\otimes\ket{1_{21}0_{22}0_{23}}
\nonumber\displaybreak[0]\\[1mm]
{}-\sin\theta_\text{I}
\ket{0_{11}0_{12}0_{13}}&\otimes\ket{0_{21}1_{22}0_{23}}
\nonumber\displaybreak[0]\\
{}-\sin\theta_\text{I}
\ket{0_{11}0_{12}0_{13}}&\otimes\ket{0_{21}0_{22}1_{23}}
\Bigr)
\nonumber\displaybreak[0]\\
&\qquad\quad
\otimes\ket{0_{31}0_{32}0_{33}},
\end{align}
and this is further converted into
\begin{widetext}
\begin{align}\label{2step}
\ket{\Psi_\text{I}}
=\frac{1}{\sqrt{3}}\,\Bigl(
\cos\theta_\text{I}
\ket{1_{11}0_{12}0_{13}}\otimes\ket{0_{21}0_{22}0_{23}}
\otimes\ket{0_{31}0_{32}0_{33}}
{}-i&\cos\theta_\text{I}
\ket{0_{11}1_{12}0_{13}}\otimes\ket{0_{21}0_{22}0_{23}}
\otimes\ket{0_{31}0_{32}0_{33}}
\nonumber\\[-1mm]
{}-i\cos\theta_\text{I}
\ket{0_{11}0_{12}1_{13}}\otimes\ket{0_{21}0_{22}0_{23}}
\otimes\ket{0_{31}0_{32}0_{33}}
&{}-i\sin\theta_\text{I}\cos\theta_\text{II}
\ket{0_{11}0_{12}0_{13}}\otimes\ket{1_{21}0_{22}0_{23}}
\otimes\ket{0_{31}0_{32}0_{33}}
\nonumber\\[1mm]
{}-\sin\theta_\text{I}\cos\theta_\text{II}
\ket{0_{11}0_{12}0_{13}}\otimes\ket{0_{21}1_{22}0_{23}}
\otimes\ket{0_{31}0_{32}0_{33}}
&{}-\sin\theta_\text{I}\cos\theta_\text{II}
\ket{0_{11}0_{12}0_{13}}\otimes\ket{0_{21}0_{22}1_{23}}
\otimes\ket{0_{31}0_{32}0_{33}}
\nonumber\\[1mm]
{}-\sin\theta_\text{I}\sin\theta_\text{II}
\ket{0_{11}0_{12}0_{13}}\otimes\ket{0_{21}0_{22}0_{23}}
\otimes\ket{1_{31}0_{32}0_{33}}
&{}+i\sin\theta_\text{I}\sin\theta_\text{II}
\ket{0_{11}0_{12}0_{13}}\otimes\ket{0_{21}0_{22}0_{23}}
\otimes\ket{0_{31}1_{32}0_{33}}
\nonumber\\
&{}+i\sin\theta_\text{I}\sin\theta_\text{II}
\ket{0_{11}0_{12}0_{13}}\otimes\ket{0_{21}0_{22}0_{23}}
\otimes\ket{0_{31}0_{32}1_{33}}
\Bigr)
\end{align}
\end{widetext}
after the second step with the couplings $g_{21,31}=g_{22,32}=g_{23,33}=g_\text{II}\neq 0$ turned on while others are off, where $\theta_\text{I/II}=g_\text{I/II}\tau_\text{I/II}/\hbar$.

Equation \eqref{1step} clearly shows that the tuning $\sin \theta_\text{I}=\cos\theta_\text{I}=1/\sqrt{2}$ realizes a \textit{one-step diffusion} of the W state up to the second row, namely from the tri-partite to a hexa-partite W state
\begin{align}\label{1step-a}
\ket{W_2}
=\frac{1}{\sqrt{6}}\,\Bigl(
\ket{1_{11}0_{12}0_{13}}&\otimes\ket{0_{21}0_{22}0_{23}}
\nonumber\\[-1mm]
{}-i\ket{0_{11}1_{12}0_{13}}&\otimes\ket{0_{21}0_{22}0_{23}}
\nonumber\\[1mm]
{}-i\ket{0_{11}0_{12}1_{13}}&\otimes\ket{0_{21}0_{22}0_{23}}
\nonumber\displaybreak[0]\\[1mm]
{}-i\ket{0_{11}0_{12}0_{13}}&\otimes\ket{1_{21}0_{22}0_{23}}
\nonumber\\[1mm]
{}-\ket{0_{11}0_{12}0_{13}}&\otimes\ket{0_{21}1_{22}0_{23}}
\nonumber\\
{}-\ket{0_{11}0_{12}0_{13}}&\otimes\ket{0_{21}0_{22}1_{23}}
\Bigr)
\otimes\ket{0_{31}0_{32}0_{33}}.
\end{align}
Or alternatively, if we select a different interaction strength so as to satisfy $\sin\theta_\text{I}=1$, another W state
\begin{equation}\label{1step-b}
\ket{w_2}
=-i
\ket{0_{11}0_{12}0_{13}}
\otimes\ket{W}_2
\otimes\ket{0_{31}0_{32}0_{33}}
\end{equation}
is established, that is, the tri-partite W state in the first row is shifted, or \textit{transferred}, to the second row after the one step.

The second step further diffuses or transfers the W state to the third row.
Indeed, Eq.\ \eqref{2step} shows that one choice $\sin\theta_\text{I}=\sqrt{2/3}$, $\cos\theta_\text{I}=1/\sqrt{3}$ and $\sin\theta_\text{II}=\cos\theta_\text{II}=1/\sqrt{2}$ yields a complete W state all over the $3\times3$ array
\begin{align}\label{2step-a}
\ket{W_3}
=\frac{1}{3}\,\Bigl(
\ket{1_{11}0_{12}0_{13}}\otimes\ket{0_{21}0_{22}0_{23}}
&\otimes\ket{0_{31}0_{32}0_{33}}
\nonumber\\[-1mm]
{}-i\ket{0_{11}1_{12}0_{13}}\otimes\ket{0_{21}0_{22}0_{23}}
&\otimes\ket{0_{31}0_{32}0_{33}}
\nonumber\\[1mm]
{}-i\ket{0_{11}0_{12}1_{13}}\otimes\ket{0_{21}0_{22}0_{23}}
&\otimes\ket{0_{31}0_{32}0_{33}}
\nonumber\\[1mm]
{}-i\ket{0_{11}0_{12}0_{13}}\otimes\ket{1_{21}0_{22}0_{23}}
&\otimes\ket{0_{31}0_{32}0_{33}}
\nonumber\\[1mm]
{}-\ket{0_{11}0_{12}0_{13}}\otimes\ket{0_{21}1_{22}0_{23}}
&\otimes\ket{0_{31}0_{32}0_{33}}
\nonumber\\[1mm]
{}-\ket{0_{11}0_{12}0_{13}}\otimes\ket{0_{21}0_{22}1_{23}}
&\otimes\ket{0_{31}0_{32}0_{33}}
\nonumber\\[1mm]
{}-\ket{0_{11}0_{12}0_{13}}\otimes\ket{0_{21}0_{22}0_{23}}
&\otimes\ket{1_{31}0_{32}0_{33}}
\nonumber\\[1mm]
{}+i\ket{0_{11}0_{12}0_{13}}\otimes\ket{0_{21}0_{22}0_{23}}
&\otimes\ket{0_{31}1_{32}0_{33}}
\nonumber\\
{}+i\ket{0_{11}0_{12}0_{13}}\otimes\ket{0_{21}0_{22}0_{23}}
&\otimes\ket{0_{31}0_{32}1_{33}}
\Bigr),
\end{align}
namely, the W state \eqref{initial} is diffused from the three qubits to the nine ones with only two steps; more generally, the W state prepared in the first row is diffused all over the $M\times N$ array via $N-1$ steps.
Another choice $\sin\theta_\text{I}=\sin\theta_\text{II}=1$ generates
\begin{align}\label{2step-b}
\ket{w_3}
=-\ket{0_{11}0_{12}0_{13}}
\otimes\ket{0_{21}0_{22}0_{23}}
\otimes\ket{W}_3
\end{align}
transferring the W state (\ref{initial}) in the first row to the third.

\section{Conclusions} \label{sec:Conclusion}
In summary, we have extended the theoretical scheme for the generation of a W entangled
state proposed in Ref.\ \onlinecite{noiPR2006} to a scheme for an $M\times N$ array of qubits.
A remarkable feature is that an entanglement realized as a W state in one row of the
array is diffused to two rows and the number of qubits involved in the W state is doubled after a single step.
The entanglement is further diffused all over the array by repeating similar processes to yield an $(M\times N)$-qubit W state.
This procedure would facilitate the generation of a large-scale multipartite W state with fewer steps, and as a result,
 would help to save time for its generation.
Furthermore, we have demonstrated the possibility of transferring the W state prepared in one row to another at will.
We have illustrated these schemes in the context of the inductively coupled flux qubits, but they are also applicable
to other systems, provided the qubit-qubit couplings are controllable.

The experimental realization of our proposal is possible, although
it has to face various technological challenges. The fabrication of
the proposed circuit is not difficult. Moreover,  to exploit tunable
qubits and flux transformers allows both to strongly reduce the
single-qubit parameter disorder  and at the same time to select
common energy gaps with a disorder of a few percent\cite{chiare
priv} and to control the duration of interqubit couplings. In
addition, state preparation can be accurately realized with
well-defined procedures including relaxation and single qubit
rotations. A delicate point concerns the system time scales against
undesired effects due to the coupling of each qubit with bosonic
baths, traceable back to the presence of many unavoidable noise
sources. For instance, the effective impedance characterizing the
dissipative electronic circuitry coupled to the single qubit
progressively degrades its coherent evolution. The consequent impact
on the single qubit decoherence rates and on the performance of a
gate of two inductively coupled qubit have been studied bringing to
light that, by carefully engineering the environmental impedances,
the bipartite systems is characterized by rates in the range
$10^{-7}$--$10^{-6}\,\text{s}$.\cite{moiij2bit} Since the observed
relaxation and decoherence times for a single flux qubit are in the
range $1$--$10\,\mu\text{s}$\cite{noritoday,chiarello2,cosmel} the
passage to a bipartite system speeds up the decoherence process.
Thus the present scheme, shortening significantly the generation
time of a multipartite W state, provides an effective way to
anticipate the occurrence of coherence loss effects. We assume that
the eigenfrequency $\omega$ of a Josephson qubit is of the order of
$10\,\text{GHz}$ and that the inductive qubit-qubit coupling
constant is of the order of
$0.5\,\text{GHz}$.\cite{noritoday,norideph} Under these conditions,
the length of a generic step (during which only a fraction of a Rabi
oscillation takes place) is  of the order of $2\,\text{ns}$. This
means that the W state \eqref{2step-a} for the $3\times3$ array is
generated approximately after $8\,\text{ns}$, which is much shorter
than $16\,\text{ns}$ required for the generation of a 9-partite W
state by the scheme discussed in Ref.\ \onlinecite{noiPR2006}.
Extending this argument to a larger array (for instance, to a
$5\times5$ array), we find that the generation of an entangled state
of $25$ qubits (which requires 8 steps)  is roughly compatible with
the currently observed relaxation times, $\approx 300\,\text{ns}$,
characterizing two coupled qubits. We wish to emphasize that this
estimation is suitable in our case, since during each step we deal
with non interacting bipartite systems of inductively coupled
qubits. If we access only one qubit at a step to entangle multiple
qubits, such a large-scale entanglement cannot be established within
such limited time.



\begin{thebibliography}{10}

\bibitem{makhlin}
Y. Makhlin, G. Sch\"on, and A. Shnirman, Rev. Mod. Phys. \textbf{73},  357
  (2001).

\bibitem{noritoday}
J.~Q. You and F. Nori, Phys. Today \textbf{58},  42  (2005).

\bibitem{DiVincenzo}
D. P. DiVincenzo, Fortschr. Phys. \textbf{48}, 771 (2000).

\bibitem{castellano}
M.~G. Castellano, F. Chiarello, R. Leoni, D. Simeone, G. Torrioli, C. Cosmelli,
  and P. Carelli, Appl. Phys. Lett. \textbf{86},  152504  (2005).

\bibitem{plourde}
B.~L.~T. Plourde, J. Zhang, K.~B. Whaley, F.~K. Wilhelm, T.~L. Robertson, T.
  Hime, S. Linzen, P.~A. Reichardt, C.-E. Wu, and J. Clarke, Phys. Rev. B
  \textbf{70},  140501(R)  (2004).

\bibitem{norideph}
Y.-X. Liu, L.~F. Wei, J.~S. Tsai, and F. Nori, Phys. Rev. Lett. \textbf{96},
  067003  (2006).

\bibitem{[3]}
E. Collin, G. Ithier, A. Aassime, P. Joyez, D. Vion, and D. Esteve, Phys. Rev.
  Lett. \textbf{93},  157005  (2004).

\bibitem{[4]}
T. Yamamoto, Y.~A. Pashkin, O. Astafiev, Y. Nakamura, and J.~S. Tsai, Nature
  (London) \textbf{425},  941  (2003).

\bibitem{[5]}
R. McDermott, R.~W. Simmonds, M. Steffen, K.~B. Cooper, K. Cicak, K.~D. Osborn,
  S. Oh, D.~P. Pappas, and J.~M. Martinis, Science \textbf{307},  1299  (2005).

\bibitem{[3a]}
A. Izmalkov, M. Grajcar, E. Il'ichev, T. Wagner, H.-G. Meyer, A.~Y. Smirnov,
  M.~H.~S. Amin, A. MaassenvandenBrink, and A.~M. Zagoskin, Phys. Rev. Lett.
  \textbf{93},  037003  (2004);
M. Grajcar, A. Izmalkov, S.~H.~W. van~der Ploeg, S. Linzen, E. Il'ichev, T.
  Wagner, U. H\"ubner, H.-G. Meyer, A. MaassenvandenBrink, S. Uchaikin, and A.~M.
  Zagoskin, Phys. Rev. B \textbf{72},  020503(R)  (2005).

\bibitem{martinis_scienze}
Y.-X. Liu, L.~F. Wei, F. Nori, Phys. Rev. B \textbf{72}, 014547
(2005); M. Steffen, M. Ansmann, R.~C. Bialczak, N. Katz, E. Lucero, R. McDermott, M. Neeley, E.~M. Weig, A.~N. Cleland, and J.~M. Martinis, Science \textbf{313}, 1423 (2006).

\bibitem{noiPR2006}

R. Migliore, K. Yuasa, H. Nakazato, and A. Messina, Phys. Rev. B \textbf{74},
  104503  (2006).

\bibitem{mooij99}
J.~E. Mooij, T.~P. Orlando, L. Levitov, L. Tian, C.~H. van~der Wal,
and S.
  Lloyd, Science \textbf{285}, 1036 (1999).


\bibitem{RSFQ}
K.~K. Likharev and V.~K. Semenov, IEEE Trans. Appl. Supercond.
\textbf{1}, 3
  (1991); V.~K. Semenov and D.~V. Averin, \textit{ibid.} \textbf{13}, 960
  (2003); D.~S. Crankshaw, J.~L. Habif, X. Zhou, T.~P. Orlando, M.~J. Feldman,
  and M.~F. Bocko, \textit{ibid.} \textbf{13}, 966 (2003); M.~G. Castellano, L.
  Gronberg, P. Carelli, F. Chiarello, C. Cosmelli, R. Leoni, S. Poletto, G.
  Torrioli, J. Hassel, and P. Helisto, 	arXiv:cond-mat/0604547 [cond-mat.supr-con] (2006).

\bibitem{chiarello2}
F. Chiarello, arXiv:cond-mat/0602464 [cond-mat.supr-con] (2006).
\bibitem{chiare priv} F. Chiarello, \textit{private communication}
(2007).
\bibitem{moiij2bit} F. K. Wilhelm, M. J. Storcz, C. H. van der Wal, C. J. P. M. Harmans, and J. E.
Mooij, in \textit{Advances in Solid State Physics}, edited by  B. Kramer (Springer-Verlag, Berlin, 2003), Vol. 43, pp. 763-780.


\bibitem{cosmel}
C. Cosmelli, P. Carelli, M.~G. Castellano, F. Chiarello, R. Leoni, B. Ruggiero,
  P. Silvestrini, and G. Torrioli, J. Supercond. \textbf{12},  773  (1999).

\end{thebibliography}

\end{document}